\documentclass[]{ifacconf}

\usepackage{graphicx}      
\usepackage{natbib}        
\usepackage{amsmath}
\usepackage{amsfonts}
\usepackage{amssymb}
\usepackage{bm}
\usepackage{mathtools}
\usepackage{xfrac}

\begin{document}

\begin{frontmatter}

\title{A dynamic system characterization of road
network node models\thanksref{footnoteinfo}} 

\thanks[footnoteinfo]{This work was supported by the
California Department of Transportation.}

\author[First]{Matthew Wright} 
\author[First]{Roberto Horowitz} 
\author[Third]{Alex A. Kurzhanskiy}

\address[First]{Mechanical Engineering Department
	and Partners for Advanced Transportation Technologies, 
   University of California, Berkeley,
   CA 94720 USA (e-mail: \{mwright, horowitz\}@berkeley.edu)}
\address[Third]{Partners for Advanced Transportation Technologies,
	University of California, Berkeley,
	CA 94720 USA (e-mail: akurzhan@berkeley.edu)}

\begin{abstract}                
The propagation of traffic congestion along roads is a commonplace nonlinear phenomenon.
When many roads are connected in a network, congestion can spill from one road to others as drivers queue to enter a congested road, creating further nonlinearities in the network dynamics.
This paper considers the node model problem, which refers to methods for solving for cross-flows when roads meet at a junction.
We present a simple hybrid dynamic system that, given a macroscopic snapshot of the roads entering and exiting a node, intuitively models the node's throughflows over time.
This dynamic system produces solutions to the node model problem that are equal to those produced by many popular node models without intuitive physical meanings.
We also show how the earlier node models can be rederived as executions of our dynamic system.
The intuitive physical description supplied by our system provides a base for control of the road junction system dynamics, as well as the emergent network dynamics.
\end{abstract}


\end{frontmatter}

\newcommand{\MI}{\mathcal{I}}
\newcommand{\MJ}{\mathcal{J}}
\newcommand{\ind}{\mathbf{1}}
\newcommand{\pt}{\tilde{p}}
\newcommand{\St}{\tilde{S}}
\newcommand{\RR}{\mathbb{R}}

\section{Introduction}
Road congestion is a major source of inefficiency in cities.
It has been estimated that, in 2014, delays due to congestion cost 7 billion hours and \$160B in the U.S. alone \citep{urban_mobility_2015}.
Modeling the nonlinear phenomenon of traffic congestion can help us understand and reduce its spread.

Often, aggregate flow behavior of many vehicles along a road is modeled similarly to a compressible fluid through a pipe.
In the simplest formulation, the movement of traffic density $\rho$ through time $t$ and lineal road direction $x$ is said to follow conservation of mass along one dimension:
\begin{equation}
\frac{\partial \rho(x,t)}{ \partial t} + \frac{ \partial f(\rho,x,t)}{\partial x} = 0, \label{eq:LWR}
\end{equation}
where $f(\rho,x,t)$ is some flux function or flow model.
The fluid-like traffic model is often called \emph{macroscopic} to contrast it with models that consider individual vehicles.

In practice, macroscopic models are simulated via finite-volume approximation in a manner similar to traditional computational fluid dynamics.
A road is broken into small cells of uniform density and Riemann problems are evaluated at each discontinuity at each simulation timestep.
Networks of roads are modeled as directed graphs.
Edges that represent individual roads are called links, and junctions where links meet are called nodes.
Typically the flow model $f(\cdot)$ on links is called the ``link model,'' and the flow model at nodes is called the ``node model.''

Traffic through a long, straight road, like other compressible flows, exhibits nonlinear phenomena such as shocks (i.e., traffic jams).
Control of traffic flows can thus require nonlinear control techniques.

When we extend from a long, straight road to a network of many roads, we create a much more complex system.
Individual links now have boundary conditions that are themselves dependent on their connected links.
A one-dimensional continuum model is not sufficient to describe road network behavior.
The node model determines how the state of an individual link affects and is affected by its connected links, their own connected links, and so on through the network.
The network-scale congestion dynamics can, in a sense, be considered as an emergent behavior rooted in the many node interactions.

Unfortunately, it is at the node model that the fluid analogy breaks down.
Unlike a physical fluid, traffic elements (i.e., vehicles) are influenced by not only physical laws, but the desires of their drivers.
A node model must take drivers' turning behavior into account.
As a result, network traffic is much more difficult to model and control than traffic along one road.

Current node models are often presented in algorithm form (i.e., presenting a series of steps for computing throughflows, see e.g. \citet{tampere11, smits_family_2015}), which can obfuscate their physical, real-world meaning.
This can have downsides: it can make it difficult to see how differences in junctions might translate to differences in node models, it can be hard to apply control methods that use differential or difference equations, and such algorithms do not have clear continuous-time meanings.

In this article, we develop a hybrid system description of a node model (recall a hybrid system is a dynamic system with both discrete and continuous states).
This model provides a more intuitive view of the abstracted real-world processes it represents, and allows for application of dynamic system control methods.
The existence of this hybrid system description of node reveals that the network-scale dynamics emerge from constituent subsystems - the nodes within.
We also show our hybrid system produces solutions equal to those of the algorithmic node models.

\section{The Node Flow Problem: Setup}

\subsection{Node models for the flow assignment problem}
\label{subsec:nodeModel}
The node model problem was first described by \citet{daganzo95a} and we reuse most of his notation.
We consider a single node and its connected links.
Let $i$ index the input links, $i \in \{1,\dots,M\}$, and $j$ index the output links, $j \in \{1,\dots,N\}$.
Individual classes of vehicles (also called commodities) are indexed by $c$, $c \in \{1,\dots,C\}$.

In the finite-volume discretization, the link model is broken into \emph{sending} and \emph{receiving} functions, which are functions of cell density $\rho$.
The sending function, $S(\rho)$, describes a cell's interactions with its downstream cell(s) and the \emph{receiving} function, $R(\rho)$, describes its interactions with upstream cell(s).
At a node, we are concerned only with the values of the sending functions of the input links and the receiving functions of the output links.
An input link's sending functions $S_i^c$ are specified per vehicle class $c$, and are equal to the number of vehicles of class $c$ that wish to exit link $i$ over the simulation timestep.
$S_i^c$ is also called the demand of class $c$ from link $i$.
An output link's receiving function $R_j$, also called link $j$'s supply, is the total number of vehicles that link $j$ can accept.\footnote{If some class $c$ takes up more supply (i.e., space) than others, its demand $S_i^c$ should be scaled appropriately.}

Drivers' behavior are encoded into \emph{split ratios}, with $\beta_{ij}^c$ denoting the portion of vehicles of class $c$ exiting link $i$ that wish to enter link $j$ ($\sum_j \beta_{ij}^c = 1$).
The quantity $S_{ij}^c \triangleq S_i^c \beta_{ij}^c$ is called the oriented demand of class $c$ from $i$ to $j$.

The role of a node model is to determine the set of flows $f_{ij}^c$ as a function of the preceding quantities.
Of importance is the behavior of a node model when there is congestion - that is, when one or more output link is unable to accept all demands directed to it.
In cases of congestion, a node model must describe how to portion the available supply $R_j$ to the demanding input links.
In this congested case, input links that wish to send vehicles to the congested link may not be able to fulfill their entire demand.
Further, if some input link $i$ wishes to send vehicles to two output links $j$ and $j'$, and $j$ becomes congested, then the congestion can be said to ``spill back'' into $i$, and reduce $i$'s ability to send vehicles to both $j$ \emph{and} $j'$.
Since drivers are assumed to be selfish, a realistic node model would take into account the disadvantage placed upon $i$'s drivers, and permit more of the supply $R_{j'}$ to be used by drivers from other links that wish to send to $j'$ exclusively, than in the counterfactual where $j$ had not been congested.
A node model's description of the (i) portioning of supplies of congested links and (ii) partial reassignment of claimed supply from links experiencing spillback was termed a ``supply constraint interaction rule'' by \citet{tampere11}.
\subsection{Mathematical Statement}

The node model problem is nearly always presented as an optimization problem.
Specifically, following \citet{wright_node_2016}, we will consider a problem partially defined as
\begin{subequations} \label{eq:optimizationProb}
\begin{align}
	&\max \sum_i \sum_j \sum_c f_{ij}^c \label{eq:optimizationMax} \displaybreak[0]\\
	&\textnormal{s.t.} \nonumber \\
	& f_{ij}^c \geq 0 \quad \forall i,j,c \displaybreak[0]\\
	& \sum_j f_{ij}^c \leq S_i^c \quad \forall i,c \label{cons:demand} \displaybreak[0]\\
	& \sum_i \sum_c f_{ij}^c \leq R_j \quad \forall j \label{cons:supply} \displaybreak[0]\\
	& \frac{f_{ij}^c}{ \sum_c f_{ij}^c} = \frac{S_{ij}^c}{ \sum_c S_{ij}^c} \quad \forall i,j,c \label{cons:comm} \displaybreak[0]\\
	& S_{ij}^c = S_i^c \beta_{ij}^c \quad \forall i,j,c \label{cons:orientedDemand} \displaybreak[0] \\ 
	& \sum_j \beta_{ij}^c = 1 \quad \forall i,c \label{cons:beta} \\
	\textnormal{Supply }&\textnormal{portioning and reallocation constraints.} \label{cons:notShownYet}
\end{align}
\end{subequations}
Constraint \eqref{cons:demand} is the demand feasibility constraint, \eqref{cons:supply} is the supply feasibility constraint, and \eqref{cons:comm} is a constraint that says each vehicle class is equally impeded by congestion.
Constraints \eqref{cons:orientedDemand} and \eqref{cons:beta}, which define the split ratios and oriented demands, were introduced above.
Constraint \eqref{cons:notShownYet}, described in Section \ref{subsec:nodeModel}, will be explained in Sections \ref{subsec:merge} and \ref{subsec:diverge}.

To solve \eqref{eq:optimizationProb}, the recent node model literature (e.g. \citet{tampere11}, \citet{smits_family_2015}, \citet{wright_node_2016}, etc.) have prescribed algorithms that ``build up'' the solution set across iterations by finding one or more of the $f_{ij}^c$ that solve \eqref{eq:optimizationProb} per iteration.
In this paper we will instead consider algorithms as dynamic systems and define a hybrid automaton whose executions solve \eqref{eq:optimizationProb} and give the same $f_{ij}^c$ as the solution algorithms (recall a hybrid automaton is a hybrid system with no control input).

\section{The dynamic system approach to the node flow problem}

Our development will parallel that of~\citet{wright_node_2016}: we will begin with the special cases of merge ($M$-to-1) and diverge (1-to-$N$) nodes, then combine the two to discuss a general merge-diverge ($M$-to-$N$) node.

The hybrid system representing the node has a discrete state space $Q$ and a continuous state space $X$.
The continuous state vector $x \in X$ is an $M \cdot N \cdot C$ real-valued vector.
Analogous to the notation for node throughflows $f_{ij}^c$, the element $x_{ij}^c$ represents the number of vehicles of class $c$ that have taken movement $i,j$.
The continuous state $x$ evolves over time according to a vector field $F$,
\begin{align}
\begin{split}
	\dot{x} &= F(q,x),
\end{split}
\end{align}
where $q \in Q$.
We also use $F_{ij}^c(q,x)$ to denote the element of $F$ corresponding to $x_{ij}^c$.
At the end of an execution, the final values of the continuous states $x_{ij}^c$ are the node throughflows $f_{ij}^c$.

\subsection{Merge node}
\label{subsec:merge}

Consider a node with $M$ input links and 1 output link.
Solving \eqref{eq:optimizationProb} in this situation is mostly straightforward: the only question is the portioning of supply among the input links.
Most contemporary node models use a supply portioning schemes that is explicit in formulation.
In these explicit formulations, each input link is assigned a priority $p_i > 0$ that represents its ability to ``claim'' a portion of the supply of the output links.
The supply that is made available to link $i$ is proportional to $p_i / \sum_{i'} p_{i'}$\footnote{
Some node models, such as the model of \citet{gibb_model_2011} and another presented in \citet{smits_family_2015} do not use explicit supply portioning schemes in the style of priorities $p_i$, but rather implicit schemes (in the sense that supply portions are defined implicitly through systems of nonlinear equations).
We will not discuss these types of node models in this article; extending hybrid system representations to those is an area of future research.
}.
When an input link $i$ does not fill its claimed proportional supply, the remainder is freed for other input links to use.

We will express this requirement as \citep{wright_node_2016}
\begin{subequations} \label{eq:mergePriority}
\begin{align}
	&\forall i',i'' \textnormal{ s.t. } f_{i'1} < S_{i'} \textnormal{ and } \, f_{i''1} < S_{i''}, \quad
	\frac{f_{i'}}{p_{i'}} = \frac{f_{i''}}{p_{i''}}, \label{eq:mergePriority1} \\
	&\textnormal{If } f_{i1} < S_i, \textnormal{ then } f_{i1} \geq \frac{p_i}{\sum_i p_i} R_1. \label{eq:mergePriority2}
\end{align}
\end{subequations}
Constraint \eqref{eq:mergePriority} looks complex because it encapsulates both the supply portioning and reassignment behavior.
Constraint \eqref{eq:mergePriority1} says that for all input links that cannot fill their whole demand, the resulting flow is proportional to their priority.
Constraint \eqref{eq:mergePriority2} says that these same congested input links may still use ``leftover'' supply - their flow is lower-bounded by their initial portioning.
A consequence is that ``leftover'' supply is also assigned proportionally to input links' priorities.
Combining \eqref{eq:mergePriority} with \eqref{eq:optimizationMax}-\eqref{cons:comm} creates the merge node problem.

We now present a hybrid automaton whose execution solves \eqref{eq:optimizationMax}-\eqref{cons:comm},\eqref{eq:mergePriority}.
\begin{defn}[Merge node hybrid system] \label{def:mergeNode}~\\
\begin{itemize}
\item Let there be $M \cdot C$ continuous states $x_{i1}^c$, each representing the number of vehicles of class $c$ that have taken movement $i,1$ through the node.
\item Let there be two discrete states, $q_\emptyset$ and $q_1$, the index representing the set of downstream links that have become congested.
The downstream link 1 is said to ``become congested'' at time $t$ if $\sum_i \sum_c x_{i1}^c(t) = R_1$.
\item Init $\subseteq Q \times X$ defines the set of permissible initial states of the system at $t=0$.
\item Dom: $Q \to X$ denotes the domain of a discrete state, which is the space of permissible continuous states while the discrete state is active.
\item $\Phi: Q \times X \to Q \times X$ is a reset relation, which defines the transitions between discrete states and the conditions for those transitions.
\end{itemize}
Then our deterministic hybrid automaton ($Q, X,$ Init, $F$, Dom, $\Phi$) is
\begin{subequations}
\begin{align}
	Q &= \{  q_\emptyset, q_1 \}, \tag{MN1} \label{merge:Q} \displaybreak[0]\\
	X &= \RR^{M \cdot C}, \tag{MN2} \label{merge:X} \displaybreak[0]\\
	\textnormal{Init} &= Q \times \{x_{i1}^c=0 \, \quad \forall i,c\}, \tag{MN3} \label{merge:init} \displaybreak[0]\\
	F_{i1}^c(q,x) &= \begin{cases}
		p_i \frac{S_{i1}^c}{\sum_c S_{i1}^c} &\textnormal{ if }
		q = q_\emptyset, \, x_{i1}^c < S_{i1}^c \\
		0 &\textnormal{ if } x_{i1}^c = S_{i1}^c \textnormal{ or } q = q_1,
	\end{cases} \tag{MN4} \label{merge:F} \displaybreak[0]\\
	\begin{split}
		\textnormal{Dom}(q_\emptyset) &=
		\left\{x: \sum_i \sum_c x_{i1}^c \leq R_1 \right\}, \textnormal{ and} \\
		\textnormal{Dom}(q_1) &= \left\{x: \sum_i \sum_c x_{i1}^c = R_1 \right\},
	\end{split} \tag{MN5} \label{merge:dom} \displaybreak[0]\\
		\Phi(q_\emptyset,x) &= (q_1, x)
			\textnormal{ if } \sum_i \sum_c x_{i1}^c = R_1.
		 \tag{MN6} \label{merge:r}
\end{align}
\end{subequations}
When $\dot{x}^c_{i1} = 0$ for all $i,c$, the execution is complete and $f_{ij}^c = x_{i1}^c$.
\end{defn}

Following \eqref{merge:init}, a system following Definition \ref{def:mergeNode} would begin with all continuous states equal to zero, and the discrete state equal to $q_\emptyset$ if $R_1 > 0$ and $q_1$ otherwise.
One can see from \eqref{merge:F} that the continuous states will increase linearly in proportion to their input link's priority.
A flow $x_{i1}^c$ stops when it is ``finished,'' i.e. when its demand is fulfilled.
Since non-fulfilled demands will still have nonzero rates of change, the ``leftover'' supply becomes proportionally available to the remaining demands.
The domain \eqref{merge:dom} and reset relation \eqref{merge:r} ensure that when $\sum_i \sum_c f_{i1}^c = R_1$, the output link is considered ``filled'' and no more vehicles can enter.

As an aside, it is reasonable to state that the rate of change $\dot{x}_{ij}^c$ in \eqref{merge:F} is independent of $S_{ij}^c$ (outside of class proportionality) because vehicles' flow rate is, in general, independent of the vehicles behind them.

\subsection{Diverge node}
\label{subsec:diverge}
Now consider a node with 1 input link and $N$ output links.
In this situation, it has been assumed (\citet{daganzo95a}, \citet{tampere11}, \citet{smits_family_2015}, etc.) that the diverging flows $f_{1j}^c$ must satisfy a ``conservation of turning ratios'' constraint,
\begin{equation}
	\frac{\sum_c f_{ij}^c}{\sum_c S_{ij}^c} = \frac{\sum_c f_{ij'}^c}{\sum_c S_{ij'}^c} \, \forall j,j'. \label{eq:fifo}
\end{equation}
The assumption behind this constraint is that, when one output link $j$ fills up, the next vehicle that wants to enter $j$ will queue at the node since the driver cannot take their desired route.
Vehicles that wish to enter other links $j'$ will be stuck behind this vehicle.
Constraint \eqref{eq:fifo}, then, is really a first-in-first-out (FIFO) constraint.

In \citet{wright_node_2016}, it was argued that this FIFO assumption may be unrealistic.
While vehicles may queue in this manner on a single-lane road, for multi-lane roads only certain lanes may queue, and traffic could still pass through other lanes.
\citet{wright_node_2016} suggested a relaxation of the FIFO constraint, and introduced \emph{mutual restriction intervals} $\bm{\eta}_{j'j}^i \subseteq [0,1]$.

The mutual restriction interval partly describes the overlapping regions of link $i$'s exit that serve links $j$ and $j'$.
Usually, this occurs with different lanes in input link $i$ allowing different sets of movements.
In a general mutual restriction interval, $\bm{\eta}_{j'j}^i = [y,z]$ means that of the lanes in input link $i$ that allow movement into link $j$, a $z-y$ portion will be blocked when link $j'$ is congested, as cars queue to enter link $j'$.
The variable $y$ denotes the leftmost extent of the blocked portion, and $z$ the rightmost extent (for example, a restriction interval for the leftmost lane of three lanes being blocked would be written $[0,\sfrac{1}{3}]$).

\begin{figure}
\centering
\includegraphics[width=8.4cm]{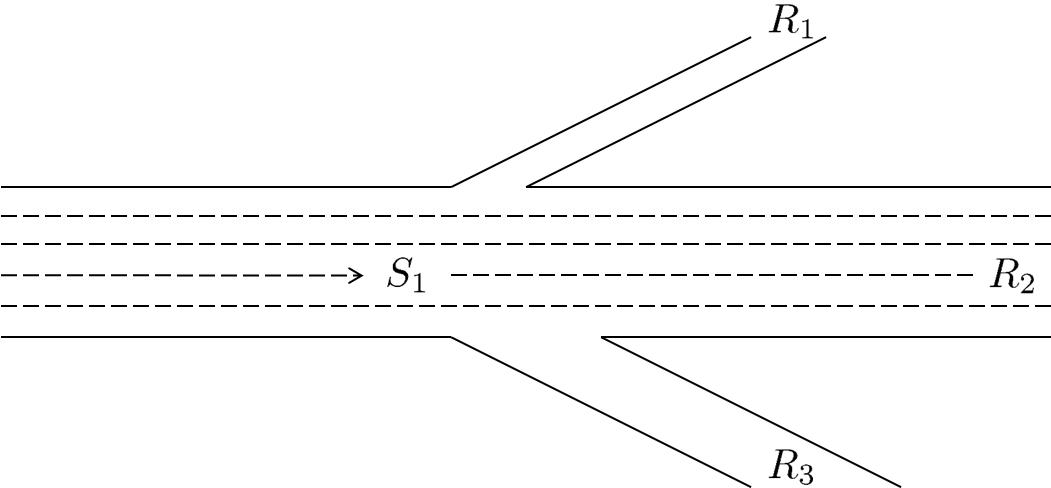}
\caption{A node with one input and three output links, where
congestion in output links~1 and~3 only partially affects
flow into output link~2, while congestion in link~2 affects
flows into output links~1 and~3 in full, and output links~1
and~3 do not affect each other.}
\label{fig:simo}
\end{figure}

Consider the example junction in Fig. \ref{fig:simo}.
If the FIFO constraint \eqref{eq:fifo} were applied, congestion in either output link $1$ or $3$ would spill back unrealistically into the mainline, blocking all entry into link 2, despite the fact that there exist lanes away from output links 1 or 3.
If, instead, we were to say that congestion in output link 1 caused queueing in the leftmost lane of the input link, and congestion in output link 3 caused queueing in the two rightmost lanes of the input link, the proper mutual restriction intervals would be, in matrix form,
\begin{equation*}
\begin{pmatrix}
	[0,1] & [0, \tfrac{1}{5}] & [0,0] \\[.5em]
	[0, 1] & [0, 1] & [0, 1] \\[.5em]
	[0, 0] & \left[\tfrac{3}{5}, 1\right] & [0, 1]
\end{pmatrix},
\end{equation*}
where $\bm{\eta}_{j'j}^1$ is the $j',j$ entry of the matrix.

\begin{figure}
\centering
\includegraphics[width=8.4cm]{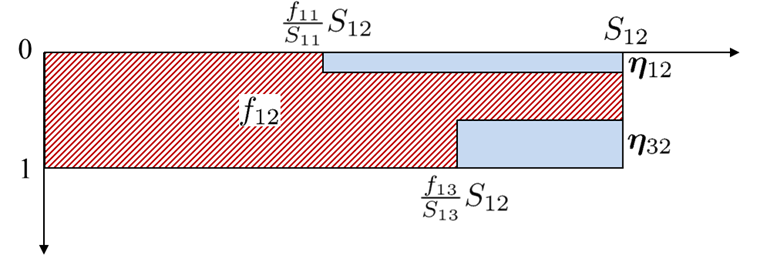}
\caption{Graphical representation of mutual restriction intervals in action.}
\label{fig:f12}
\end{figure}

In \citet{wright_node_2016}, the application of the mutual restriction intervals was elaborated in a graphical manner.
Fig. \ref{fig:f12} shows an example of calculating $f_{12}$ for the junction of Fig. \ref{fig:simo}.
This figure describes a situation where output link 1 and output link 3 run out of supply before all of $S_{12}$ is satisfied.
In particular, $\bm{\eta}_{12}$ affects more of $S_{12}$ than does $\bm{\eta}_{32}$.
The most intuitive interpretation is that link 1 fills before link 3.
The intervals $ [\frac{f_{j'j'}}{S_{j'j'}} S_{j'j}, \, S_{j'j}]$ along the horizontal axis indicate the extent to which $\bm{\eta}_{j'j}$ is applied.
In this formulation, we replace the FIFO constraint \eqref{eq:fifo} with a relaxed FIFO constraint in the form
\begin{equation}
	f_{1j}^c \leq S_{1j}^c - \mathcal{A} \left( \bigcup_{j' \neq j} 
		\left\{ \bm{\eta}_{j'j}^1 \times \left[ \frac{f_{1j'}}{S_{1j'}} S_{1j}, \, S_{1j} \right] \right\}
		\right), \label{eq:rectangleRelaxedFifo}
\end{equation}
where $\mathcal{A}(\cdot)$ denotes the area of the two-dimensional object.

The interpretation of Fig. \ref{fig:f12}, where link 1 becomes filled ``before'' link 3, suggests that we can also solve the diverge node problem  \eqref{eq:optimizationMax}-\eqref{cons:comm}, \eqref{eq:rectangleRelaxedFifo} with a hybrid system that activates mutual restriction intervals during its executions.

By definition, the number of vehicles that are trying to leave the input link over the simulation timestep is $\sum_c S_1^c$, and those trying to move to link $j$ is $\sum_c S_{1j}^c$.
This value assumes use of all possible lanes over the entire simulation timestep.
Since we are taking $p_1$ to be the rate at which the input link sends vehicles and claims downstream supply, the length of the simulation timestep is $(\sum_c S_1^c)/p_1 \triangleq T_1$.
If relaxed FIFO is in effect, some portion of the time period $[0,T_1]$ may have vehicles exiting the input link at a lower rate.
Specifically, when link $j'$ becomes filled, movement $1,j$ will have only $1 - | \bm{\eta}_{j'j}^1 |$ (where $| \bm{\eta}_{j'j}^1 |$ represents the length of the interval) of its maximum possible lanes available, so the flow along movement $i,j$ will proceed at $1 - | \bm{\eta}_{j'j}^1 |$ times its maximum possible rate.
Should another link become congested ($j''$, say), then the flow rate will be $1 - |~\bm{\eta}_{j'j}^1~\cup~\bm{\eta}_{j''j}^1 |$ times the maximum.

Finally, recall we defined $p_1$ as the rate at which input link 1 sent vehicles out of it.
In the merge node case, vehicles leaving a link only had one destination. but here there are $N$ possible destinations.
The exiting flow rate $p_1$ is thus split among the destinations.
Since we already know the portions that the vehicles themselves are split among destinations, portioning the priority is simple.
We introduce $p_{1j}$, the ``oriented priority'' as a shorthand:
\begin{equation}
	p_{1j} \triangleq \frac{\sum_c S_{1j}^c}{\sum_c S_1^c} p_1. \label{eq:orientedPriority}
\end{equation}

\begin{defn}[Diverge node hybrid system]
	\label{def:divergeNode}
	~\\ \begin{itemize}
		\item The notation for $x_{1j}^c$ and $\dot{x}_{1j}^c$ parallel Definition \ref{def:mergeNode}.
		\item Our discrete states are still indexed by the output link(s) with zero remaining supply. Let $\mathcal{J}$ be the set of all output links.
		\item Let $j^\ast$ denote a particular output link that has had its supply filled.
		\item The hybrid system execution begins at time $t=0$.
	\end{itemize}
	Our hybrid automaton ($Q, X,$ Init, $F$, Dom, $\Phi$) is
	\begin{subequations}
	\begin{align}
		Q &= \{  q_\nu \}, \, \nu \in 2^\mathcal{J}, \tag{DN1} \label{diverge:Q} \displaybreak[0]\\
		X &= \RR^{N \cdot C}, \tag{DN2} \label{diverge:X} \displaybreak[0]\\
		\textnormal{Init} &= Q \times \{x_{1j}^c=0 \, \quad \forall j,c\}, \tag{DN3} \label{diverge:init} \displaybreak[0]\\
		F_{1j}^c(q,x) &= \begin{cases}
			\begin{aligned}[c]
			p_{1j} & \frac{ S_{1j}^c}{ \sum_c S_{1j}^c} \\
			\cdot& \bigg(1 -\big| \bigcup_{ \mathclap{ j' \in \nu } } \bm{\eta}_{j'j}^1 \big| \bigg)
			\end{aligned}
			& \begin{aligned}[c] &\textnormal{ if }  \, x_{1j}^c < S_{1j}^c \\
								 &\textnormal{ and } \, t < T_1
			\end{aligned} \\[1em]
			0 &\textnormal{ otherwise}, 
		\end{cases} \tag{DN4} \label{diverge:F} \displaybreak[0]\\
			\textnormal{Dom}(q_\nu) &=
			\left\{ \begin{aligned} 
				x: &\sum_c x_{1j}^c = R_j \quad \forall j \in \nu \textnormal{ and} \\
					&\sum_c x_{1j}^c \leq R_j \quad \forall j \notin \nu
				\end{aligned} \right\}, \tag{DN5} \label{diverge:dom} \displaybreak[0]\\
		\Phi(q_\nu,x) &= 
		\begin{aligned}[t]
			(q_{\nu'}, x) &\textnormal{ if } \sum_c x_{1j^\ast}^c = R_{j^\ast}, \\
				&\textnormal{ where } \nu' = \nu \cup j^\ast.
		\end{aligned} \tag{DN6} \label{diverge:r}
	\end{align}
	\end{subequations}
	Recall that $2^\mathcal{J}$ as in \eqref{diverge:Q} denotes the power set of $\mathcal{J}$. 
	When $\dot{x}^c_{1j} = 0$ for all $j,c$, the execution is complete.
\end{defn}

Examining the definition, one sees that the system switches among the discrete modes as output links' supplies exhaust, similarly to in the merge case.
As links' supplies fill, \eqref{diverge:r} adds them to the set $\nu$ and the flow rates are attenuated in \eqref{diverge:F} as described above.
Equation \eqref{diverge:F} shows that the flow rate of any movement will become zero in any of three conditions: (i) its demand is exhausted, (ii) the input link's ``maximum time'' $T_1$ expires, or (iii) the downstream supply is exhausted.
To see how (iii) occurs, note that $\bm{\eta}_{jj}^1 = [0,1]$ for all $j$ (i.e., the diagonal entires in the matrix above), so the relaxed FIFO factor will equal zero in \eqref{diverge:F}.

\subsection{General node formulation}
The general node problem is \eqref{eq:optimizationMax}-\eqref{cons:comm}, \eqref{eq:mergePriority}, \eqref{eq:rectangleRelaxedFifo}.
Now that we have seen the merge and diverge cases, combining them into a general node formulation is not too difficult.
The only items to keep in mind involve the mutual restriction intervals.
The first is that there can be mutual restriction intervals per input, $\bm{\eta}_{j'j}^i$.
Secondly, in generalizing the use of mutual restriction intervals to multiple inputs, we say that the intervals $\bm{\eta}_{j'j}^i$ should not be enforced on movements $i,j$ if there are no vehicles that want to take the movement $i,j'$.
The reason for this is simple: if there is no remaining demand for the movement $i,j'$, there will be no queue to block lanes, even if link $j'$ has become full.

\begin{defn}[General node hybrid system]
	~\\ \begin{itemize}
		\item The notation for $x_{ij}^c$, $\dot{x}_{ij}^c$, $j^\ast$, and $\mathcal{J}$ parallel the earlier definitions.
		\item The hybrid system execution begins at time $t=0$.
		\item The upper time limit for each input link generalizes that in the diverge node: $T_i = S_i/p_i$.
	\end{itemize}
	Our hybrid automaton ($Q, X,$ Init, $F$, Dom, $\Phi$) is
	\begin{subequations}
	\begin{align}
		Q &= \{  q_\nu \}, \, \nu \in 2^\mathcal{J}, \tag{GN1} \label{general:Q} \displaybreak[0]\\
		X &= \RR^{M \cdot N \cdot C}, \tag{GN2} \label{general:X} \displaybreak[0]\\
		\textnormal{Init} &= Q \times \{x_{ij}^c=0 \, \quad \forall i,j,c\}, \tag{GN3} \label{general:init} \displaybreak[0]\\
		F_{ij}^c(q,x) &= \begin{cases}
			\begin{aligned}[c]
			p_{ij} & \frac{ S_{ij}^c}{\sum_c S_{ij}^c} \\
			\cdot& \bigg(1 - \big| \bigcup_{ \mathclap{
			\substack{j' \in \nu,\\	\exists \, c: \, x_{ij'}^c < S_{ij'}^c }} }
			\bm{\eta}_{j'j}^i \big| \bigg)
			\end{aligned}
			& \begin{aligned}[c] &\textnormal{ if }  \, x_{ij}^c < S_{ij}^c \\
							 &\textnormal{ and } \, t < T_i
			\end{aligned} \\[1em]
			0 &\textnormal{ otherwise},
		\end{cases} \tag{GN4} \label{general:F} \displaybreak[0]\\
			\textnormal{Dom}(q_\nu) &=
			\left\{ \begin{aligned} 
				x: &\sum_i \sum_c x_{ij}^c = R_j \quad \hspace{-.25em} \forall j \in \nu \textnormal{ and} \\
					&\sum_i \sum_c x_{ij}^c \leq R_j \quad \hspace{-.25em} \forall j \notin \nu
				\end{aligned} \right\}, \tag{GN5} \label{general:dom} \displaybreak[0]\\
		\Phi(q_\nu,x) &= 
		\begin{aligned}[t]
			(q_{\nu'}, x) &\textnormal{ if } \sum_i \sum_c x_{ij^\ast}^c = R_{j^\ast}, \\
				&\textnormal{ where } \nu' = \nu \cup j^\ast.
		\end{aligned} \tag{GN6} \label{general:r}
	\end{align}
	\end{subequations}
	When $\dot{x}^c_{ij} = 0$ for all $i,j,c$, the execution is complete and $f_{ij}^c = x_{ij}^c$.
	\label{def:generalNode}
\end{defn}

Note that $\nu$ at the beginning of the execution may not necessarily be empty; there may be some output links with zero available supply.

Examining the system, one sees that an execution will progress in much the same manner as the diverge system in Definition \ref{def:divergeNode}.
Moving from the diverge node to the general node, one must notice that each state's $x_{ij}^c$ continuous dynamics are now linearly dependent on the input link priority $p_i$. 
This means that, in situations where (relaxed) FIFO is not in action, vehicles will use downstream supply in proportion to the input link priority, as in the merge node.

It is also useful to note how the continuous dynamics change when one or more mutual restriction intervals are in action.
In this situation, the term in the parenthesis in~\eqref{general:F} will be between 0 and 1 for at least one movement $i,j$.
In this case, one or more of the oriented priorities $p_{ij}$ \eqref{eq:orientedPriority} will be decreased, and their sum will not add up to $p_i$.
This mechanism is the method through which relaxed FIFO blocks traffic - blocked lanes cannot claim their (proportional to $p_i$) downstream supply.

\section{Applications and Extensions}

\subsection{Event-triggered evaluation}

Evaluating continuous-time or hybrid systems typically involves forward integration of the differential equation(s) with fixed or varying step sizes.
However, in the case of the systems presented in this article, evaluation can be performed in a much simpler manner.
This is due to the particular dynamics of the system - since the continuous-time dynamics and the condition for discrete mode switching are very simple, the time that the next mode switch will occur can be found in closed form.
In the case of the general node, \eqref{general:dom} and \eqref{general:r} say that a mode switch where link $j$ enters $\nu$ will occur when
\begin{equation}
	\sum_i \sum_c x_{ij}^c = R_{j}. \label{eq:modeSwitch1}
\end{equation}
Say we are currently at time $t_0$. 
Combining \eqref{eq:modeSwitch1} with \eqref{general:F}, we can find the time that the mode switch occurs, which we denote $t_{j}$.
\begin{equation}
	R_{j} = \sum_i \sum_c x_{ij}^c(t_0) + \int_{t_0}^{t_j} \sum_i \sum_c \dot{x}_{ij}^c dt. \label{eq:modeSwitch2}
\end{equation}
Solving the integral in \eqref{eq:modeSwitch2},
\begin{align}
	\begin{split}
	& \int_{t_0}^{t_j} \sum_i \sum_c \dot{x}_{ij}^c dt \\
	&= \int_{t_0}^{t_j} \sum_i \sum_c p_{ij} \frac{ S_{ij}^c}{ \sum_c S_{ij}^c}
	\bigg(1 - \big| \bigcup_{ \mathclap{
			\substack{j' \in \nu,\\	\exists \, c: \, x_{ij'}^c < S_{ij'}^c }} }
			\bm{\eta}_{j'j}^i \big| \bigg) dt
	\end{split} \nonumber \displaybreak[0]\\
	&= (t_j - t_0) \sum_i p_{ij}
		\bigg(1 - \big| \bigcup_{ \mathclap{
				\substack{j' \in \nu,\\	\exists \, c: \, x_{ij'}^c < S_{ij'}^c }} }
				\bm{\eta}_{j'j}^i \big| \bigg). \label{eq:modeSwitch3}
\end{align}
Then, plugging \eqref{eq:modeSwitch3} into \eqref{eq:modeSwitch2},
\begin{equation}
	t_j = t_0 + \frac{ R_j - \sum_i \sum_c x_{ij}^c(t_0)}{ \sum_i p_{ij} \bigg( 1 - \big| \bigcup\limits_{ \mathclap{
					\substack{j' \in \nu,\\	\exists \, c: \, x_{ij'}^c < S_{ij'}^c }} }
					 \bm{\eta}_{j'j}^i \big| \bigg) }. \label{eq:modeSwitch4}
\end{equation}
This value can be computed for each output link $j$.
Then, the smallest $t_j$ will be the first link to fill and join $\nu$.
We had used $j^\ast$ for this output link, so let $t_{j^\ast} \triangleq \min t_j$.
However, one of the input links may have its time limit $T_i$ expire.
This would also change the dynamics, as it stops sending vehicles at that time.

Therefore, evaluation of the system trajectory beginning from $t_0$ can be done by (i) evaluating \eqref{eq:modeSwitch4} for each output link, (ii) identifying $t_j^\ast$, and (iii) checking whether any of the time limits $T_i$ occur before $t_{j^\ast}$.
This is an event-triggered simulation: it is only necessary to determine when the next event will occur.
The equations for $\dot{x}_{ij}^c$ over $[t_0, \min (\{T_i\}, t_{j^\ast})]$ can then be evaluated in closed form under $q_\nu$.

\subsection{Previous node models as event-triggered simulations}
The discussion in the previous section brings us to one of the major results of this article.
Modern node models existing in the literature, such as in \citet{tampere11} and the references therein, two of \citet{smits_family_2015}, and \citet{wright_node_2016} present algorithms for solving the node flow problem that yield solutions that are identical to the steady state solution of the hybrid automaton described here.
Common among these are (i) evaluations similar to \eqref{eq:modeSwitch4} to determine which of the output links is most-demanded and (ii) comparison of the most-demanded supply (i.e. $R_{j^\ast}$) with input demands to determine whether some input links may be fully served.

In particular, the special case of \eqref{eq:modeSwitch4} where $t_0=0$ and $x_{ij}^c(t_0)=0$ may be compared with formulas for a term called $a_j$ in both \citet{tampere11} and \citet{wright_node_2016} ((26) in the former, (3.45) in the latter); they are identical.
The context of $a_j$ in both is as a ``reduction factor'' that determines what portion of demand is served, i.e. $f_{ij}^c = a_j S_{ij}^c$.
In light of this article's discussion, it is clear that this reduction factor is just proportional to the length of time that $\dot{x}_{ij}^c$ is integrated.
It is easy to also see that our ``time limit'' condition $T_i < t_{j^\ast}$ is equivalent to the conditions in both of these references that determine if an input link $i$ is in freeflow (in the language of the references, demand-constrained).

These node models, then, are just event-triggered executions of this article's presented hybrid systems.
This elucidates a physical meaning of the flows computed by these previous models.

\subsection{Non-discretized network simulation}
A particularly interesting potential application for these node models is macroscopic simulation of road networks without finite-volume discretization.
It is known that, for certain choices of the flux function $f(\cdot)$, there exist closed-form solutions $\rho(x,t)$ for the conservation equation \eqref{eq:LWR} along a link for all $x$ and $t$, given an initial condition $\rho(x,0)$ and upstream and downstream flow boundary conditions for all $t$.

The lack of a continuous-time formulation for junction flows has prevented application of these closed-form solutions to road networks.
However, the dynamic system node models presented here may be applied: the demands $S_{ij}^c$, supplies $R_j$, and flows $f_{ij}^c$ may be made time-varying.
Network simulation could then be done by evaluating the closed-form link solutions for demands and supplies, and using the node models to forward-integrate the resulting link inflows and outflows.

\subsection{Node supply constraints}
Some authors have suggested adding node supply constraint(s) to \eqref{eq:optimizationProb} describing situations where the node becomes jammed, or some ``shared resource'' used by some or all movements exhausts (e.g., a green time constraint in a signalized intersection considered by \citet{tampere11}).
A straightforward extension of our system might add these supply constraints to the domain and reset relation constructions, so the discrete state changes when the node has transmitted as many vehicles as it can handle, similar to the behavior when an output link supply $R_j$ fills.

\section{Conclusion}
This article presented hybrid system formulations of processes used to model traffic flow through a road junction.
We also showed how node models existing in the literature that had been presented in unintuitive algorithmic forms can be re-derived as executions of these hybrid systems.
This development means that the continuous-time dynamics stated here are implicit but unstated in previous node models' assumptions.
Future research can take this into account, as the continuous-time physical meaning of quantities like input link priorities and restriction intervals are apparent.
This can lead to development of more accurate and realistic node models for traffic control applications, as well as continuous-time simulation of networks. 


\bibliography{traffic}
\bibliographystyle{ifacconf}

\end{document}